\documentstyle[aps,epsfig,multicol]{revtex}
\renewcommand{\narrowtext}{\begin{multicols}{2}\global\columnwidth20.5pc}
\renewcommand{\widetext}{\end{multicols}\global\columnwidth42.5pc}

\newcommand{\wideequationbegin}{
\widetext
\noindent
\begin{picture}(3.375,0)
  \put(0,0){\line(1,0){3.375}}
  \put(3.375,0){\line(0,1){0.08}}
\end{picture}}

\newcommand{\wideequationend}{
\hfill
\begin{picture}(3.375,0)
  \put(0,0){\line(1,0){3.375}}
  \put(0,0){\line(0,-1){0.08}}
\end{picture}
\narrowtext}

\begin{document}
\title{Conductance enhancement in quantum point\\ contact--semiconductor--superconductor devices}

\draft 

\author{Niels Asger Mortensen and Antti-Pekka Jauho} 

\address{Mikroelektronik Centret, Technical University of Denmark,
  Building 345 east, DK-2800 Lyngby, Denmark}

\author{Karsten Flensberg}

\address{\O rsted Laboratory, Niels Bohr Institute, University of
  Copenhagen, Universitetsparken 5, DK-2100 Copenhagen \O, Denmark}

\author{Henning Schomerus}

\address{Instituut--Lorentz, Universiteit Leiden, P.O. Box 9506, 2300
  RA Leiden, The Netherlands} 

\maketitle

\begin{abstract}
  We present numerical calculations of the conductance of an interface
  between a phase-coherent two-dimensional electron gas and a
  superconductor with a quantum point contact in the normal region.
  Using a scattering matrix approach we reconsider the geometry of De
  Raedt, Michielsen, and Klapwijk [Phys. Rev. B, {\bf 50}, 631 (1994)]
  which was studied within the time-dependent Bogoliubov--de Gennes
  formalism. We find that the factor-of-two enhancement of the
  conductance $G_{\scriptscriptstyle\rm NS}$ compared to the normal
  state conductance $G_{\scriptscriptstyle\rm N}$ for ideal interfaces
  may be suppressed for interfaces with a quantum point contact with
  only a few propagating modes. The suppression is found to depend
  strongly on the position of the Fermi level. We also study the
  suppression due to a barrier at the interface and find an anomalous
  behavior caused by quasiparticle interference. Finally, we consider
  the limit of sequential tunneling and find a suppression of the
  factor-of-two enhancement which may explain the absence of
  conductance enhancement in experiments on metal--superconductor
  structures.
\end{abstract}
 

\narrowtext

\section{Introduction}

Charge transport through a normal conductor--superconductor (NS)
interface is accompanied by a conversion of quasiparticle current to a
supercurrent. In the Andreev reflection, by which the conversion
occurs, an electron-like quasiparticle in the normal conductor (with
an excitation energy lower than the energy gap of the superconductor)
incident on the NS interface is retroreflected into a hole-like
quasiparticle (with reversal of its momentum and its energy relative
to the Fermi level) and a Cooper pair is added to the condensate of
the superconductor \cite{ANDREEV64}. For an ideal NS interface, the
signature of Cooper pair transport and the Andreev scattering is a
doubling of the conductance compared to the normal state conductance.

A theoretical framework for studies of the scattering at NS interfaces
is provided by the Bogoliubov--de Gennes (BdG) formalism
\cite{DEGENNES66} where the scattering states are eigenfunctions of
the BdG equation

\begin{equation}
\left(\begin{array}{cc}\hat{\cal H}({\bf r})& \Delta({\bf r})\\
    \Delta^*({\bf r})& -\hat{\cal H}^*({\bf r}) \end{array}\right)
\left(\begin{array}{c}u({\bf r})\\v({\bf r})\end{array}\right)
=E\left(\begin{array}{c}u({\bf r})\\v({\bf
    r})\end{array}\right)\, ,
\end{equation}
which is a Schr\"{o}dinger-like equation in electron-hole space (Nambu
space). Here $\hat{\cal H}({\bf r})$ is the single-particle
Hamiltonian, $\Delta({\bf r})$ is the pairing potential of the
superconductor, $E$ is the excitation energy, and $u({\bf r})$ and
$v({\bf r})$ are wave functions of electron-like and hole-like
quasiparticles.

The technological possibility of studying the interface between a
two-dimensional electron gas (2DEG) in a semiconductor heterostructure
and a superconductor experimentally, has provided a playground for
investigating the interplay between the Andreev reflection and the
mesoscopic effects seen in mesoscopic semiconductor structures
\cite{BEENAKKER91c,IMRY}. In the recent years the technological
efforts have revealed a variety of new mesoscopic phenomena, see Refs.
\cite{IMRY,KLAPWIJK94,BEENAKKER95,LAMBERT98} and references therein.
One class of the studied devices are the quantum point contact (QPC)
2DEG-S and S-2DEG-S devices with the QPC in the normal region. The dc
Josephson effect and the quantization of the critical current in QPC
S-2DEG-S junctions have been studied extensively both experimentally
by e.g.  Takayanagi and co-workers \cite{TAKAYANAGI} and theoretically
by e.g.  Beenakker and van Houten \cite{BEENAKKER91a,VANHOUTEN91},
Beenakker \cite{BEENAKKER91b}, and Furusaki, Takayanagi, and Tsukada
\cite{FURUSAKI}.

The linear-response conducting properties of QPC 2DEG-S structures
have been studied by several groups. In the analytical work of
Beenakker \cite{BEENAKKER92} a ballistic normal region with a QPC
modeled by a saddle-point potential was considered. The effect of
elastic impurity scattering was considered numerically by Takagaki and
Takayanagi \cite{TAKAGAKI96} who considered a disordered region
between a narrow-wide (NW) constriction and the superconductor. Both
of these studies of the conductance were based on a scattering matrix
(S--matrix) approach and the BdG formalism. In the numerical
simulations of De Raedt, Michielsen, and Klapwijk \cite{RAEDT94},
based on the time-dependent BdG equation, a wide-narrow-wide (WNW)
constriction was considered. Here, the aim was to study the
electron-hole conversion efficiency and the robustness of the
back-focusing phenomena of the Andreev reflection.

One of the properties of the QPC is that most transmission eigenvalues
are either close to zero or unity. For an ideal QPC 2DEG-S interface,
the Andreev reflection will therefore give rise to a factor-of-two
enhancement of the conductance $G_{\scriptscriptstyle\rm NS}$ compared
to the normal state conductance $G_{\scriptscriptstyle\rm N}$
\cite{VANHOUTEN91,BEENAKKER92}, which is quantized in units of
$2e^2/h$ \cite{VANWEES88}. However, as pointed out by van Houten and
Beenakker \cite{VANHOUTEN91}, deviations from the simple factor-of-two
enhancement should be expected when the position of the Fermi level
does not correspond to a conductance plateau. The presence of impurity
scattering in the normal region and/or interface roughness will also
suppress the doubling of the conductance \cite{TAKAGAKI96}.

\begin{figure}
\begin{center}
\epsfig{file=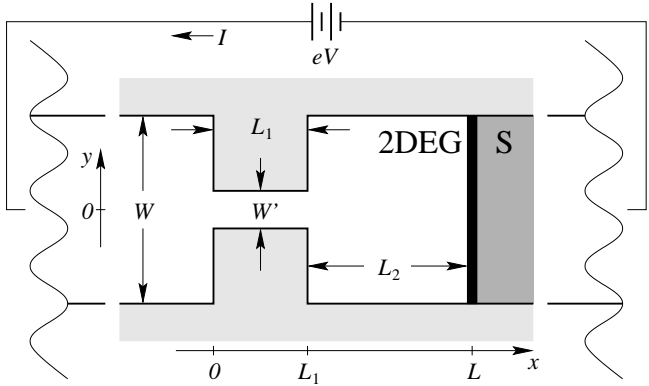, width=0.99\columnwidth}
\end{center}
\caption[]{Geometry of a WNW 2DEG-S junction with a hard-wall confining
  potential and a barrier at the 2DEG-S interface.}
\label{FIG1}
\end{figure}

Using an S--matrix approach, we study the linear-response regime of a
phase-coherent ballistic QPC 2DEG-S system where the QPC is modeled by
a WNW constrictions with a hard-wall confining potential, see Fig.
\ref{FIG1}. We report new results for the device studied by De
Raedt \emph{et al.} \cite{RAEDT94} which had a relative width $W/W'
=1.7\,{\rm \mu m}/(1.6\times 335\,{\rm \AA})\simeq 31.72$, an aspect
ratio $L_1/W' = 5/1.6$, and a relative length $L_2/W'=20/1.6$.
Applying the S--matrix formalism instead of the computationally more
complicated time-dependent BdG formalism, we are able to study a
larger part of the parameter-space where we also consider a barrier
(with a normalized barrier strength $Z$) at the NS interface. We focus
on the regime with only a few propagating modes in the QPC.
In this regime the transmission eigenvalues of the QPC depend strongly
on the actual position of the Fermi level. Even for an ideal interface
this gives rise to a strong suppression of the conductance for certain
positions of the Fermi level as predicted by van Houten and Beenakker
\cite{VANHOUTEN91} and subsequently seen in the work of Beenakker
\cite[Fig. 1]{BEENAKKER92}. In the presence of a barrier at the
interface, the QPC gives rise to an enhanced tunneling through the
barrier (compared to the case without a QPC) as in the case of
reflectionless tunneling effect of diffusive junctions
\cite{BEENAKKER92,VANWEES92}.

In the sequential tunneling limit the conductance can be found by
considering the QPC and the interface as two series-connected
resistive regions and in the limit $W\gg W'$ the enhancement of the
conductance compared to the normal state conductance vanishes even for
ideal NS interfaces. This may be an explanation for the unexpectedly
low conductance enhancement in the experimental results of Benistant
\emph{et al.} \cite{BENISTANT85} on Ag-Pb interfaces where the current
is injected into the Ag crystal through a point contact.

The text is organized as follows: In Section II the S--matrix
formalism is introduced, in Section III we formulate our model, in
Section IV the scattering scheme of the considered geometry is
presented, and in Sections V we present results of several
applications of our scattering scheme. Finally, in Section VI
discussion and conclusions are given.

\section{Scattering matrix formalism}

The scattering approach to coherent dc transport in superconducting
hybrids follows closely the scattering theory developed for
non-superconducting mesoscopic structures, see e.g. the text-book by
Datta \cite{DATTA95}.

For an ideal NS interface, the interface acts as a phase-conjugating
mirror within the Andreev approximation \cite{ANDREEV64} and the rigid
boundary condition for the pairing potential
\begin{equation}
\Delta({\bf r}) = \Delta_0 e^{i\varphi}\Theta(x-L)\,,
\end{equation}
where $\Delta_0$ is the BCS energy gap \cite{BCS57}, $\varphi$ is the
phase of the pairing potential, $\Theta(x)$ is a Heaviside function,
and $L=L_1+L_2$ is the length of the normal region (see Fig. \ref{FIG1}).

In the linear-response regime in zero magnetic field, Beenakker
\cite{BEENAKKER92} found that the conductance $G\equiv\partial I/\partial V$
is given by
\begin{eqnarray}
G_{\scriptscriptstyle\rm NS}
&=& \frac{4e^2}{h} {\rm Tr}\,
\left(t t^\dagger \left[2\hat{1} -tt^\dagger\right]^{-1} \right)^2\nonumber\\
&=& \frac{4e^2}{h}\sum_{n=1}^{N} \frac{T_n^2}{\left(2-T_n\right)^2},
\label{BEENAKKER}
\end{eqnarray}
which, in contrast to the Landauer formula \cite{FISHERLEE81},
\begin{equation}
G_{\scriptscriptstyle\rm N}
=  \frac{2e^2}{h}{\rm Tr}\, t t^\dagger 
= \frac{2e^2}{h}\sum_{n=1}^{N} T_n,
\label{LANDAUER}
\end{equation}
is a non-linear function of the transmission eigenvalues $T_n$
($n=1,2,\ldots,N$) of $t t^\dagger$. Here $t$ is the $N\times N$
transmission matrix of the normal region, $N$ being the number of
propagating modes.

Equation (\ref{BEENAKKER}) holds for an arbitrary disorder potential
and is a multi-channel generalization of a conductance formula first
obtained by Blonder, Tinkham and Klapwijk \cite{BTK82} who considered
a delta function potential as a model for the interface barrier
potential. The computational advantage of Eq.  (\ref{BEENAKKER}) over
the time-dependent BdG approach of De Raedt \emph{et al.}
\cite{RAEDT94} is that we only need to consider the time-independent
Schr\"{o}dinger equation with a potential which describes the disorder
in the normal region, so that we can use the techniques developed for
quantum transport in normal conducting mesoscopic structures.

For more details on e.g. finite bias and/or temperature, see Lesovik,
Fauch\`ere and Blatter \cite{LESOVIK97}, Lesovik and Blatter
\cite{LESOVIK98}, and the reviews of Beenakker \cite{BEENAKKER95}, and
Lambert and Raimondi \cite{LAMBERT98}.

\section{Model}

We describe the geometry of Fig. \ref{FIG1} by the Hamiltonian

\begin{equation}
\label{Hamiltonian}
\hat{\cal H}({\bf r}) =
- \frac{\hbar^2}{2m}\hat{\bf \nabla}
+ V_{\scriptscriptstyle\rm \delta}({\bf r})
+ V_{\scriptscriptstyle\rm c}({\bf r}) -\mu\,,
\end{equation}
where $\mu$ is the chemical potential. The barrier potential is given
a Dirac delta function potential with strength $H$ \cite{BTK82}:
\begin{equation}
V_{\scriptscriptstyle\rm \delta}({\bf r})=H\delta(x-L)\,,
\end{equation}
and the transverse motion is limited by a hard-wall confining potential

\begin{equation}
V_{\scriptscriptstyle\rm c}({\bf r}) =\left\{\begin{array}{ccc}
0 &,& |y|< {\cal W}(x)/2\\
\infty &,& |y|\geq {\cal W}(x)/2
\end{array}\right.\,,
\end{equation}
where the width ${\cal W}(x)$ defines the WNW constriction and is
given by
\begin{equation}
{\cal W}(x)=\left\{\begin{array}{ccc}
W&,&x<0\\
W'&,&0\leq x \leq L_1\\
W&,&x>L_1
\end{array}\right.\,.
\end{equation}
The scattering states can be constructed as linear combinations of
the eigenstates of the Schr\"{o}dinger equation.

\section{Scattering scheme}

\begin{figure}
\begin{center}
\epsfig{file=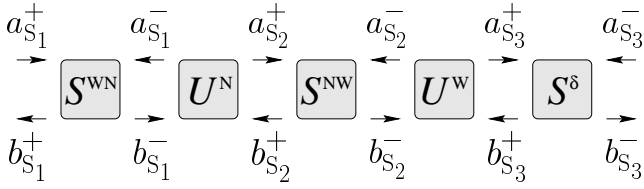, width=0.99\columnwidth}
\end{center}
\caption[]{Scattering scheme appropriate for the normal region of the
  geometry shown in Fig. \ref{FIG1}.}
\label{FIG2}
\end{figure}

In the following subsections, we consider the S--matrices of a system
with the Hamiltonian in Eq. (\ref{Hamiltonian}) relevant for the
geometry shown in Fig. \ref{FIG1}. The S-matrix $S$ of a scattering
region relates the incident current amplitudes
$a_{\scriptscriptstyle\rm S}^\pm$ to the outgoing current amplitudes
$b_{\scriptscriptstyle\rm S}^\pm$. For a scattering region with two
leads, $S$ is a $2\times 2$ block-matrix with sub-matrices $S_{11}$,
$S_{12}$, $S_{21}$, and $S_{22}$, where the diagonal and off-diagonal
sub-matrices are reflection and transmission matrices, respectively.
The appropriate scattering scheme for three scattering regions (the WN
and the NW constrictions, and the interface barrier potential,
respectively) connected by ballistic conductors is shown in Fig.
\ref{FIG2}.

The WN constriction is described by the S--matrix
$S^{\scriptscriptstyle\rm WN}$
\begin{equation}
\left(\begin{array}{c}
b_{\scriptscriptstyle\rm S_1}^+\\
b_{\scriptscriptstyle\rm S_1}^-
\end{array}\right)
=S^{\scriptscriptstyle\rm WN} 
\left(\begin{array}{c}
a_{\scriptscriptstyle\rm S_1}^+\\ 
a_{\scriptscriptstyle\rm S_1}^-
\end{array}\right),
\end{equation}
the narrow region of length $L_1$ by the propagation-matrix
$U^{\scriptscriptstyle\rm N}$
\begin{equation}
\left(\begin{array}{c}
a_{\scriptscriptstyle\rm S_1}^-\\ 
a_{\scriptscriptstyle\rm S_2}^+
\end{array}\right)=U^{\scriptscriptstyle\rm N} 
\left(\begin{array}{c}
b_{\scriptscriptstyle\rm S_1}^-\\ 
b_{\scriptscriptstyle\rm S_2}^+
\end{array}\right),
\end{equation}
the NW constriction by the S--matrix $S^{\scriptscriptstyle\rm NW}$
\begin{equation}
\left(\begin{array}{c}
b_{\scriptscriptstyle\rm S_2}^+\\ 
b_{\scriptscriptstyle\rm S_2}^-
\end{array}\right)=S^{\scriptscriptstyle\rm NW} 
\left(\begin{array}{c}
a_{\scriptscriptstyle\rm S_2}^+\\ 
a_{\scriptscriptstyle\rm S_2}^-
\end{array}\right),
\end{equation}
the wide region of length $L_2$ by the propagation-matrix
$U^{\scriptscriptstyle\rm W}$
\begin{equation}
\left(\begin{array}{c}
a_{\scriptscriptstyle\rm S_2}^-\\ 
a_{\scriptscriptstyle\rm S_3}^+
\end{array}\right)=U^{\scriptscriptstyle\rm W} 
\left(\begin{array}{c}
b_{\scriptscriptstyle\rm S_2}^-\\ 
b_{\scriptscriptstyle\rm S_3}^+
\end{array}\right),
\end{equation}
and the delta function barrier by the S--matrix
$S^{\scriptscriptstyle\rm \delta}$
 \begin{equation}
\left(\begin{array}{c}
b_{\scriptscriptstyle\rm S_3}^+\\ 
b_{\scriptscriptstyle\rm S_3}^-
\end{array}\right)=S^{\scriptscriptstyle\rm \delta} 
\left(\begin{array}{c}
a_{\scriptscriptstyle\rm S_3}^+\\ 
a_{\scriptscriptstyle\rm S_3}^-
\end{array}\right).
\end{equation}

To apply Eqs. (\ref{BEENAKKER}) and (\ref{LANDAUER}) we need to
calculate the composite transmission matrix $t=S_{21}$ which is a
sub-matrix of the composite S--matrix $S\equiv
S^{\scriptscriptstyle\rm WN}\otimes U^{\scriptscriptstyle\rm N}\otimes
S^{\scriptscriptstyle\rm NW}\otimes U^{\scriptscriptstyle\rm W}\otimes
S^{\scriptscriptstyle\rm \delta}$ relating the outgoing current
amplitudes to the incoming current amplitudes,
\begin{equation}
\left(\begin{array}{c}
b_{\scriptscriptstyle\rm S_1}^+\\ 
b_{\scriptscriptstyle\rm S_3}^-
\end{array}\right)=S \left(\begin{array}{c}
a_{\scriptscriptstyle\rm S_1}^+\\ 
a_{\scriptscriptstyle\rm S_3}^-
\end{array}\right)\,,\, 
S=\left(\begin{array}{cc}
r& t'\\
t& r'
\end{array}\right).
\end{equation}
The meaning of the symbol $\otimes$ is found by eliminating the
internal current amplitudes \cite{DATTA95}. As a final result we find
the transmission matrix
\wideequationbegin
\begin{eqnarray}
t&=&S_{21}^{\scriptscriptstyle\rm \delta}
\bigg[
\hat{1}-U_{21}^{\scriptscriptstyle\rm W}
\left\{
S_{22}^{\scriptscriptstyle\rm NW} 
+ S_{21}^{\scriptscriptstyle\rm NW}
\left[
\hat{1}-U_{21}^{\scriptscriptstyle\rm N}
S_{22}^{\scriptscriptstyle\rm WN}
U_{12}^{\scriptscriptstyle\rm N}
S_{11}^{\scriptscriptstyle\rm NW} 
\right]^{-1} 
U_{21}^{\scriptscriptstyle\rm N}
S_{22}^{\scriptscriptstyle\rm WN}
U_{12}^{\scriptscriptstyle\rm N}
S_{12}^{\scriptscriptstyle\rm NW}
\right\} 
U_{12}^{\scriptscriptstyle\rm W}
S_{11}^{\scriptscriptstyle\rm \delta}
\bigg]^{-1} \nonumber\\
&\quad &\times U_{21}^{\scriptscriptstyle\rm W} 
S_{21}^{\scriptscriptstyle\rm NW}
\left[
\hat{1}-U_{21}^{\scriptscriptstyle\rm N}
S_{22}^{\scriptscriptstyle\rm WN}
U_{12}^{\scriptscriptstyle\rm N}
S_{11}^{\scriptscriptstyle\rm NW}
\right]^{-1}
U_{21}^{\scriptscriptstyle\rm N} 
S_{21}^{\scriptscriptstyle\rm WN}\,.
\label{t_composite}
\end{eqnarray}
\narrowtext

\subsection{Quantum point contact}

We consider a QPC which we model by a WNW constriction defined by a
hard-wall confining potential, see Fig. \ref{FIG1}. This geometry has
been considered by Szafer and Stone \cite{SZAFER89} and Weisshaar,
Lary, Goodnick and Tripathi \cite{WEISSHAAR91} in the context of
conductance quantization of the QPC in a 2DEG, and recently by
Kassubek, Stafford and Grabert \cite{KASSUBEK99} in the context of
conducting and mechanical properties of ideal two and
three-dimensional metallic nanowires. We follow Kassubek \emph{et al.}
\cite{KASSUBEK99} and calculate the composite S--matrix
$S^{\scriptscriptstyle\rm WNW} = S^{\scriptscriptstyle\rm WN}\otimes
U^{\scriptscriptstyle\rm N}\otimes S^{\scriptscriptstyle\rm NW}$. In
zero magnetic field, where all S--matrices satisfy $S=S^T$, the
individual S--matrices are given by

\begin{eqnarray}
S^{\scriptscriptstyle\rm WN} 
&=&\left(\begin{array}{cc}
r_{\scriptscriptstyle\rm NW}'  & t_{\scriptscriptstyle\rm NW}\\
t_{\scriptscriptstyle\rm NW}^T & r_{\scriptscriptstyle\rm NW}
\end{array}\right)\,,\\
U^{\scriptscriptstyle\rm N} 
&=&\left(\begin{array}{cc}\hat{0} & X^{\scriptscriptstyle\rm N}\\
X^{\scriptscriptstyle\rm N}       & \hat{0} 
\end{array}\right)
\label{U}\,,\\
S^{\scriptscriptstyle\rm NW} 
&=&\left(\begin{array}{cc}
r_{\scriptscriptstyle\rm NW} & t_{\scriptscriptstyle\rm NW}^T\\
t_{\scriptscriptstyle\rm NW} & r_{\scriptscriptstyle\rm NW}'
\end{array}\right)\,,
\end{eqnarray}
where $X_{nn'}^{\scriptscriptstyle\rm N}=\delta_{nn'}\exp\left(ik_n
  L_1\right)$ describes the narrow region with free propagation of
propagating modes and an exponential decay of evanescent modes. Here
$k_n=k_{\rm F} \sqrt{1-\left(n\pi/k_{\rm F}W'\right)^2}$ is the
longitudinal wave vector of mode $n$ in the narrow region. The
S-matrices of the WN and NW constrictions are related through an
exchange of leads.

By elimination of the internal current amplitudes we find the
composite transmission matrix

\begin{equation}
\label{WNW_S_Matrix}
S^{\scriptscriptstyle\rm WNW} 
= \left(\begin{array}{cc}
r_{\scriptscriptstyle\rm WNW} & t_{\scriptscriptstyle\rm WNW}\\ 
t_{\scriptscriptstyle\rm WNW} & r_{\scriptscriptstyle\rm WNW}
\end{array}\right)\,,
\end{equation}
where
\begin{eqnarray}
r_{\scriptscriptstyle\rm WNW}
&=& r_{\scriptscriptstyle\rm NW}' \nonumber\\
&\quad& +t_{\scriptscriptstyle\rm NW}
\left[
\hat{1}-
\left(
X^{\scriptscriptstyle\rm N} 
r_{\scriptscriptstyle\rm NW} 
\right)^2 
\right]^{-1}
X^{\scriptscriptstyle\rm N} 
r_{\scriptscriptstyle\rm NW} 
X^{\scriptscriptstyle\rm N}  
t_{\scriptscriptstyle\rm NW}^T
\label{rWNW}\,,\\
t_{\scriptscriptstyle\rm WNW}
&=& t_{\scriptscriptstyle\rm NW} 
\left[
\hat{1}-
\left(
X^{\scriptscriptstyle\rm N} 
r_{\scriptscriptstyle\rm NW}
\right)^2 
\right]^{-1} 
X^{\scriptscriptstyle\rm N}
t_{\scriptscriptstyle\rm NW}^T. 
\label{tWNW}
\end{eqnarray}

The S--matrix of the NW constriction can be found from a matching of
scattering states which are eigenstates of the Schr\"{o}dinger
equation with the Hamiltonian in Eq. (\ref{Hamiltonian}) where we only
consider the part of the potential which sets up the NW constriction. From a matching of scattering states we find that

\begin{eqnarray}
r_{\scriptscriptstyle\rm NW} 
&=& \left(
\hat{1}+ \varrho\varrho^T
\right)^{-1}
\left(
\hat{1}- \varrho\varrho^T
\right) 
\label{rNW}\,,\\
t_{\scriptscriptstyle\rm NW} 
&=& 2\varrho^T 
\left(
\hat{1}+ \varrho\varrho^T
\right)^{-1}
\label{tNW}\,,\\
r_{\scriptscriptstyle\rm NW}' 
&=& \left(
\varrho^T\varrho +\hat{1}
\right)^{-1}
\left(
\varrho^T\varrho -\hat{1}
\right)
\label{rNW'}\,,
\end{eqnarray}
where the elements of the $\varrho$-matrix can be written as
$\varrho_{nw}= \sqrt{K_w/k_n}\left< \phi_n\right|\left.\Phi_w \right>$
where $\left< \phi_n\right|\left.\Phi_w
\right>=\int_{-\infty}^{\infty}{\rm d}y\,\phi_n(y)\Phi_w(y)$ is an
overlap between transverse wave functions of mode $n$ in the narrow
region and mode $w$ in the wide region. Here $K_w=k_{\rm F}
\sqrt{1-\left(w\pi/\eta k_{\rm F}W'\right)^2}$ is the longitudinal
wave vector of mode $w$ in the wide region and $\eta\equiv W/W'$ is
the relative width of the constriction.

The overlap can be calculated analytically since its elements consist
of overlaps between transverse wave functions $\phi_n$ and $\Phi_w$
which are either two Sine or two Cosine functions (the overlap between
a Sine function and a Cosine function or vice versa is zero due to the
odd and even character of the two functions). From the overlap-matrix
we get the following elements of the $\varrho$-matrix
\wideequationbegin
\begin{equation}
\varrho_{nw}=
\delta_{\scriptscriptstyle {\cal P}(n), {\cal P}(w)}
\left( 
\frac{ 
\left(
\frac{k_{\rm F} W'}{\pi}\right)^{2}
-\left(
\frac{n}{\eta}
\right)^2}
{\left(
\frac{k_{\rm F} W'}{\pi}
\right)^{2}-n^2}
\right)^{1/4}
\times \left\{ 
\begin{array}{lcl}
\delta_{\scriptscriptstyle{\cal P}(n),1} (-1)^{(n+2)/2} 
\times
\frac{4n\eta^{3/2}\sin(w\pi/2\eta)}{\pi\left(n^2\eta^2-w^2\right)}
&,& n\eta\neq w\\
\delta_{\scriptscriptstyle{\cal P}(n),-1}  (-1)^{(n-1)/2} 
\times
\frac{4n\eta^{3/2}\cos(w\pi/2\eta)}{\pi\left(n^2\eta^2-w^2\right)}
&,&n\eta\neq w\\
\eta^{-1/2}&,& n\eta = w
\end{array}
\right.\,,
\end{equation}
\wideequationend
\noindent where the parity ${\cal P}(j)$ of $j$ is ${\cal P}(j)\equiv 1$ if $j$
is even and ${\cal P}(j)\equiv -1$ if $j$ is odd.

In the numerical evaluation of Eqs. (\ref{rNW})-(\ref{rNW'}) and Eqs.
(\ref{rWNW}) and (\ref{tWNW}) it is crucial to let the number of modes
in the narrow and wide regions extend over both propagating modes and
evanescent modes. After all matrix inversions are performed, the
reflection and transmission matrices are projected onto the
propagating modes. In practice, numerical convergence of the
reflection and transmission matrices is found for a finite cut-off in
the number of evanescent modes. For the considered device, the number
of evanescent modes is roughly ten times the number of propagating
modes in the wide region corresponding to $1000\times 1000$ matrices.

In the limit $W\gg W'$, Szafer and Stone \cite{SZAFER89} employed a
mean-field approximation for the overlap $\left<
  \phi_n\right|\left.\Phi_w \right>$ in which mode $n$ of the narrow
region couples uniformly to only a band of modes (with the same parity
as mode $n$) in the wide region within one level spacing so that the
elements of the $\varrho$-matrix take the form
\wideequationbegin
\begin{equation}\label{MFA}
\varrho_{nw}
\approx\delta_{\scriptscriptstyle {\cal P}(n), {\cal P}(w)}
\left( 
\frac{ 
\left(
\frac{k_{\rm F} W'}{\pi}
\right)^{2}
-\left(
\frac{n}{\eta}
\right)^2}
{\left(
\frac{k_{\rm F} W'}{\pi}
\right)^{2}
-n^2}
\right)^{1/4}
\times \left\{ \begin{array}{ccc}
\eta^{-1/2}&,&(n-1)\eta\leq w <(n+1)\eta\\
0&,& {\rm otherwise}\end{array}\right. .
\end{equation} 
\wideequationend
\noindent Within this approximation, there is no mode mixing and $tt^\dagger$
becomes diagonal with the transmission eigenvalues along the diagonal.
This approximation was found to capture the results of an exact
numerical calculation \cite{SZAFER89} when used with the Landauer
formula, Eq. (\ref{LANDAUER}), which is linear in the transmission
eigenvalues $T_n$. However, for an NS interface, the conductance
formula, Eq. (\ref{BEENAKKER}), is non-linear in $T_n$ which also
makes off-diagonal components in $tt^\dagger$ important. As we shall
see (in Fig. \ref{FIG3}), the mean-field approximation cannot
reproduce the results of an exact numerical calculation of
$G_{\scriptscriptstyle\rm NS}$ in the same nice way as for
$G_{\scriptscriptstyle\rm N}$.

\subsection{Wide region}
The wide region $L_1<x<L$ is described similarly to the narrow
region by
\begin{equation}
U^{\scriptscriptstyle\rm W} 
=\left(\begin{array}{cc}
\hat{0} &X^{\scriptscriptstyle\rm W}\\
X^{\scriptscriptstyle\rm W}&\hat{0} 
\end{array}\right)\,,
\end{equation}
where $X_{ww'}^{\scriptscriptstyle\rm W}=\delta_{ww'}\exp\left(iK_w
  L_2\right)$ describes both the free propagating modes and the
exponential decay of the evanescent modes. Here $K_w$ is the
introduced longitudinal wave vector of mode $w$ in the wide region.

\subsection{Interface barrier potential}

We consider an NS interface of width $W$ with a barrier which we model
by a Dirac delta function potential, following Blonder, Tinkham and
Klapwijk \cite{BTK82}. The S--matrix elements for the delta-function
potential is found from a matching of scattering states which are
eigenstates of the Schr\"{o}dinger equation with the Hamiltonian in
Eq. (\ref{Hamiltonian}) where we only consider the part of the
potential consisting of the barrier at the interface. In zero magnetic
field one finds the symmetric result
\begin{equation}
S^{\scriptscriptstyle\rm \delta}
=\left(\begin{array}{cc}
r_{\scriptscriptstyle\rm \delta} & t_{\scriptscriptstyle\rm \delta}\\
t_{\scriptscriptstyle\rm \delta}&r_{\scriptscriptstyle\rm \delta}
\end{array}\right)\,,
\end{equation}
with
\begin{eqnarray}
\left(t_{\scriptscriptstyle\rm\delta}\right)_{ww'}
&=&\delta_{ww'} \frac{1}{1 + i Z/\cos\theta_w }\,,\\
\left(r_{\scriptscriptstyle\rm\delta}\right)_{ww'}
&=&\delta_{ww'} \frac{-iZ/\cos\theta_w }{1 + iZ/\cos\theta_w  }\,,
\end{eqnarray}
where the normalized barrier strength is given by $Z\equiv H/\hbar
v_{\rm F}$ and $\cos\theta_w\equiv K_w/k_{\rm
  F}=\sqrt{1-\left(w\pi/k_{\rm F}W\right)^2}$. The results differ from
those of a one-dimensional calculation \cite{BTK82} since we have
taken the parallel degree of freedom into account. However, if we
introduce an angle dependent effective barrier strength
$Z_{\scriptscriptstyle\rm eff}(\theta_w)=Z/\cos\theta_w$
\cite{KUPKA97,MORTENSEN99}, the transmission and reflection amplitudes
can formally be written in the one-dimensional form of Ref.
\cite{BTK82}. The transmission eigenvalues of
$t_{\scriptscriptstyle\rm \delta}t_{\scriptscriptstyle\rm
  \delta}^\dagger$ are given by $T_w^{\scriptscriptstyle\rm
  \delta}=\left(1+Z_{\scriptscriptstyle\rm
    eff}^2(\theta_w)\right)^{-1}$ in contrast to the mode-independent
result $T_w^{\scriptscriptstyle\rm \delta}=\left(1+Z^2\right)^{-1}$ of
a one-dimensional calculation \cite{BTK82}.

\section{Results}

\subsection{Phase-coherent junction with ideal interface}

\begin{figure}
\begin{center}
\epsfig{file=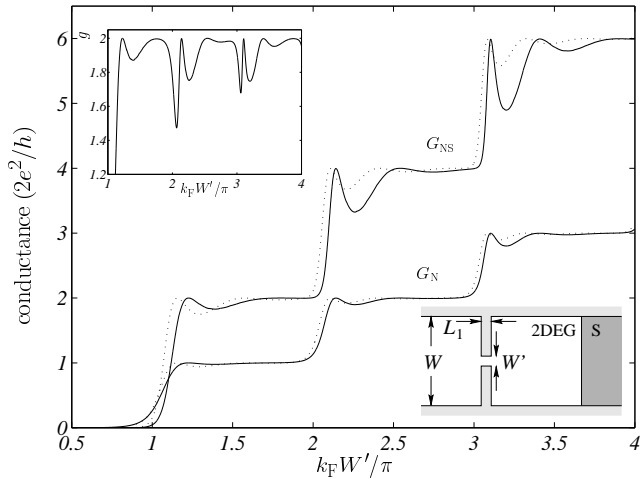, width=0.99\columnwidth}
\end{center}
\caption[]{Conductance $G_{\scriptscriptstyle\rm NS}$ and normal state
  conductance $G_{\scriptscriptstyle\rm N}$ of a coherent WNW 2DEG-S
  junction as a function of $k_{\rm F}W'/\pi$. The constriction has an
  aspect ratio $L_1/W'=1$ and a relative width $W/W'=31.72$. The full
  lines are for a numerical calculation of $t_{\scriptscriptstyle\rm
    WNW}$ and the dotted lines are results within the mean-field
  approximation of Szafer and Stone \cite{SZAFER89}. The upper insert
  shows the normalized conductance $g\equiv G_{\scriptscriptstyle\rm
    NS}/G_{\scriptscriptstyle\rm N}$ also as function of $k_{\rm
    F}W'/\pi$.}
\label{FIG3}
\end{figure}

For the case of coherent transport through an ideal 2DEG-S interface
with a WNW constriction in the normal region, see lower insert of Fig.
\ref{FIG3}, the conductance $G_{\scriptscriptstyle\rm NS}$ and the
normal state conductance $G_{\scriptscriptstyle\rm N}$ can be found
from Eqs. (\ref{BEENAKKER}) and (\ref{LANDAUER}) with the transmission
matrix $t=t_{\scriptscriptstyle\rm WNW}$.

Figure \ref{FIG3} shows the conductance as a function of $k_{\rm
  F}W'/\pi$ based on a numerical calculation of
$t_{\scriptscriptstyle\rm WNW}$ (full lines) and the mean-field
approximation (dashed line) for a WNW constriction with an aspect
ratio $L_1/W'=1$ and a relative width $W/W'=31.72$. The conductance
$G_{\scriptscriptstyle\rm NS}$ is seen to be approximately quantized
in units of $4e^2/h$ which is twice the unit of conductance for the
normal state conductance $G_{\scriptscriptstyle\rm N}$. However, just
above the thresholds ($k_{\rm F}W'/\pi=1,2,3,\ldots$), oscillations
due to resonances in the narrow region of the constriction are
observed. In the normal state result, these resonances are small but
in contrast to the Landauer formula, $G_{\scriptscriptstyle\rm NS}$ is
not linear in the transmission eigenvalues and this makes the
resonances much more pronounced compared to those in the normal state
conductance. Another signature of the non-linearity of
$G_{\scriptscriptstyle\rm NS}$ and the importance of off-diagonal
transmission, is that the mean-field approximation is in good
agreement with the numerical calculation for $G_{\scriptscriptstyle\rm
  N}$ whereas it has difficulties in accounting for
$G_{\scriptscriptstyle\rm NS}$. The sharpness of the resonances is to
a certain extent due to the wide-narrow-wide constriction, and is
suppressed in experiments with split-gate-defined constrictions.
However, as shown by the simulations of Maa\o, Zozulenko, and Hauge
\cite{MAAO94} resonance effects do persist even for more smooth
connections of the narrow region to the 2DEG reservoirs.

The normalized conductance $g\equiv G_{\scriptscriptstyle\rm
  NS}/G_{\scriptscriptstyle\rm N}$, shown in the upper insert, is two
on the conductance plateaus but for certain ``mode-fillings'' of the
constriction it is strongly suppressed and for $k_{\rm F}W'/\pi \sim
2$ (two propagating modes) we get $g\sim 1.5$. This effect, which
occurs at the onset of new modes, was also seen in the calculations of
Beenakker \cite[Fig. 1]{BEENAKKER92}.  As the number of modes
increases, these dips vanish and the normalized conductance approaches
its ideal value of two. The reason is simple: suppose the constriction
has $N$ propagating modes, then the $N-1$ of them will have a
transmission of order unity and only a single mode (corresponding to
the mode with the highest transverse energy) will have transmission
different from unity. As $N$ increases, the effect of the single mode
with transmission different from unity becomes negligible for the
normalized conductance and from Eqs. (\ref{BEENAKKER}) and
(\ref{LANDAUER}) it follows that $\lim_{N\rightarrow \infty}g= 2$.

Since the quasiparticle propagation is coherent and the Andreev
scattering is the only back-scattering mechanism, the phase
conjugation between electron-like and hole-like quasiparticles makes
the conductance $G_{\scriptscriptstyle\rm NS}$ independent of the
separation $L_2$ of the constriction and the interface. If evanescent
modes in this region were also taken into account the results would
depend weakly on $L_2$ as it was found in the simulations of De Raedt
\emph{et al.}  \cite{RAEDT94} and our results should be compared with
their results in the large $L_2$-limit. As we shall see below,
interfaces with a finite barrier (and thereby normal scattering at the
interface) lead to size-quantization and thereby resonances which will
depend on $L_2$.

The back-focusing phenomenon of the Andreev reflected quasiparticles
and the lowering of the normalized conductance due to a QPC in the
normal region was studied by De Raedt, Michielsen and Klapwijk
\cite{RAEDT94} by solving the time-dependent BdG equation fully
numerically. In their wave propagation simulations, the QPC is also
modeled by a WNW constriction with a relative width $W/W'=1.7\,{\rm
  \mu m}/(1.6\times 335\,{\rm \AA})\simeq 31.72$, an aspect ratio
$L_1/W' = 5/1.6$ and a relative length $L_2/W'=20/1.6$. For the
particular ``mode-filling'' $k_{\rm F} W'/\pi = 3.2$, they find a
normalized conductance $g=1.87 < 2$, but the dependence on the
``mode-filling'' was not studied in detail.

\begin{figure}
\begin{center}
\epsfig{file=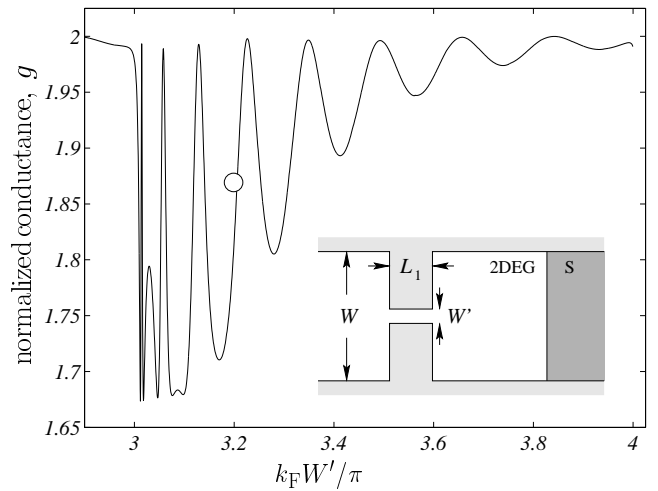, width=0.99\columnwidth}
\end{center}
\caption[]{Normalized conductance $g\equiv G_{\scriptscriptstyle\rm
    NS}/G_{\scriptscriptstyle\rm N}$ of a coherent WNW 2DEG-S junction
  as a function of $k_{\rm F}W'/\pi$. The constriction has an aspect
  ratio $L_1/W'=5/1.6$ and a relative width $W/W'=31.72$. The curve is
  based on a numerical calculation of $t_{\scriptscriptstyle\rm WNW}$.
  The data-point ($\circ$) corresponds to the numerical result
  $(k_{\rm F}W'/\pi;g)=(3.2;1.87)$ of De Raedt \emph{et al.}
  \cite[TABLE I]{RAEDT94}.}
\label{FIG4}
\end{figure}

In Fig. \ref{FIG4} we present a calculation of $g$ as a function of
$k_{\rm F}W'/\pi$ for this specific geometry. The result of De Raedt
\emph{et al.} ($\circ$) is reproduced but in general the normalized
conductance is seen to have many resonances caused by the high aspect
ratio of the constriction. In the range $3 <k_{\rm F}W'/\pi <4$, the
normalized conductance can be anything in the range $1.655 < g \leq 2$
depending on the position of the Fermi level and though De Raedt
\emph{et al.} \cite{RAEDT94} found the back-focusing phenomena of the
Andreev reflection to be very robust with respect to changes of the
device-parameters, the normalized conductance itself certainly depends
strongly on the position of the Fermi level. The reason is that only
those quasiparticles which enter the region between the constriction
and the interface can be Andreev reflected and thus contribute to the
conductance enhancement compared to the normal state conductance.

\subsection{Phase-coherent junction with barrier at interface}

\begin{figure}
\begin{center}
\epsfig{file=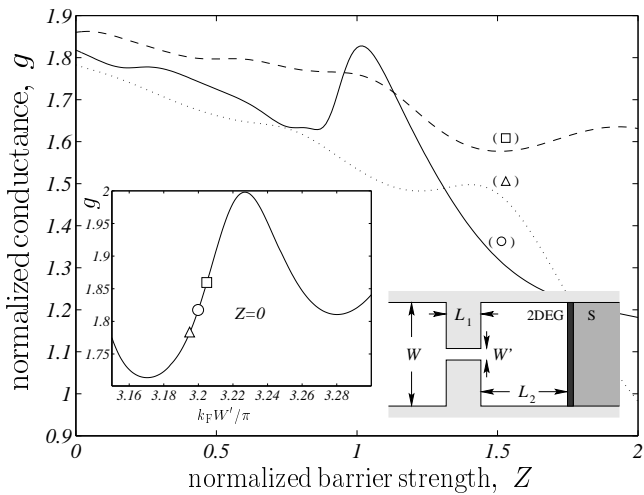, width=0.99\columnwidth}
\end{center}
\caption[]{Normalized conductance $g\equiv G_{\scriptscriptstyle\rm
    NS}/G_{\scriptscriptstyle\rm N}$ of a coherent WNW 2DEG-S junction
  with a barrier as a function of the normalized barrier strength $Z$
  for $k_{\rm F}W'/\pi = 3.195$ ($\triangle$), $k_{\rm F}W'/\pi = 3.2$
  ($\circ$), and $k_{\rm F}W'/\pi = 3.205$ ($\Box$). The lower left
  insert shows the normalized conductance $g$ as a function of $k_{\rm
    F}W'/\pi$ for $Z=0$. The constriction has an aspect ratio
  $L_1/W'=5/1.6$, a relative width $W/W'=31.72$, and the cavity has
  a relative length $L_2/W'=20/1.6$.}
\label{FIG5}
\end{figure}

We next consider coherent transport through an NS interface with a
barrier at the interface and a WNW constriction at a distance $L_2$
from the interface, see lower right insert of Fig. \ref{FIG5}. The
conductance $G_{\scriptscriptstyle\rm NS}$ and the normal state
conductance $G_{\scriptscriptstyle\rm N}$ are found from Eqs.
(\ref{BEENAKKER}) and (\ref{LANDAUER}) with the transmission matrix in
Eq. (\ref{t_composite}).

In Fig. \ref{FIG5} we present a calculation of the normalized
conductance $g$ as a function of the normalized barrier strength $Z$
for the device considered by De Raedt \emph{et al.}  \cite{RAEDT94}.
For the position of the Fermi level ($\circ$) considered by De Raedt
\emph{et al.}, the normalized conductance is only weakly suppressed
(compared to a system without a QPC, see e.g. \cite{MORTENSEN99}) for
low barrier scattering ($Z<1$) and only for a very high barrier
strength ($Z>2$) the normalized conductance approaches the cross-over
from an excess conductance ($g>1$) to a deficit conductance ($g<1$).
The effect of the barrier for $Z<1$ is very similar to the
reflectionless tunneling behavior in diffusively disordered junctions
\cite{BEENAKKER92,VANWEES92} where the net result is as if tunneling
through the barrier is reflectionless. In the case of a QPC instead of
a diffusive region there is a weak dependence on the barrier strength
and the tunneling is not perfectly reflectionless.

An interesting feature is the non-monotonic behavior of $g$ as a
function of $Z$. For $Z\rightarrow \infty$, the normalized conductance
of course vanishes, but in some regions it increases with an
increasing barrier strength [curve ($\circ$)] and for $Z\simeq 1$ it
has the same value as for $Z=0$. This is purely an effect of size
quantization in the cavity between the QPC and the barrier which
enters the conducting properties because of the fully coherent
propagation of electrons and holes. However, changing e.g.  the
position of the Fermi level slightly, [curves ($\triangle$) and
($\Box$)], changes the quantitative behavior although the
overall suppression of $g$ with increasing $Z$ is maintained.

\subsection{Incoherent junction}

In junctions where the propagation in the cavity between the QPC and
the NS interface is incoherent, the so-called sequential tunneling
regime, the QPC and the NS interface can be considered as two
series-connected resistive regions \cite{BUTTIKER}. This means that
\begin{eqnarray}
G_{\scriptscriptstyle\rm QPC-NS}
&=&\left(
G_{\scriptscriptstyle\rm QPC}^{-1}+G_{\scriptscriptstyle\rm NS}^{-1}
\right)^{-1}\,,\\
G_{\scriptscriptstyle\rm QPC-N}
&=&\left(
G_{\scriptscriptstyle\rm QPC}^{-1}+G_{\scriptscriptstyle\rm N}^{-1}
\right)^{-1}\,,
\end{eqnarray}
where $G_{\scriptscriptstyle\rm QPC}$ and $G_{\scriptscriptstyle\rm
  N}$ are found from Eq. (\ref{LANDAUER}) with
$t=t_{\scriptscriptstyle\rm QPC}$ and $t=t_{\scriptscriptstyle\rm
  \delta}$, respectively, and $G_{\scriptscriptstyle\rm NS}$ from Eq.
(\ref{BEENAKKER}) with $t=t_{\scriptscriptstyle\rm \delta}$.  The
normalized conductance can be written as

\begin{equation}
g
\equiv \frac{G_{\scriptscriptstyle\rm QPC-NS}}
{G_{\scriptscriptstyle\rm QPC-N}}
=\frac{G_{\scriptscriptstyle\rm NS}}
{G_{\scriptscriptstyle\rm N}} \times 
\frac{G_{\scriptscriptstyle\rm QPC}+ G_{\scriptscriptstyle\rm N}}
{G_{\scriptscriptstyle\rm QPC}+ G_{\scriptscriptstyle\rm NS}}\,,
\end{equation}
and for $W\gg W'$ the major contribution to the resistance comes from
the QPC, i.e. $G_{\scriptscriptstyle\rm QPC} \ll
\left(G_{\scriptscriptstyle\rm N},\, G_{\scriptscriptstyle\rm
    NS}\right)$. This means that the enhancement of
$G_{\scriptscriptstyle\rm NS}$ compared to $G_{\scriptscriptstyle\rm
  N}$ has a negligible effect on the total conductance so that the
normalized conductance approaches $g\sim 1$.

\begin{figure}
\begin{center}
\epsfig{file=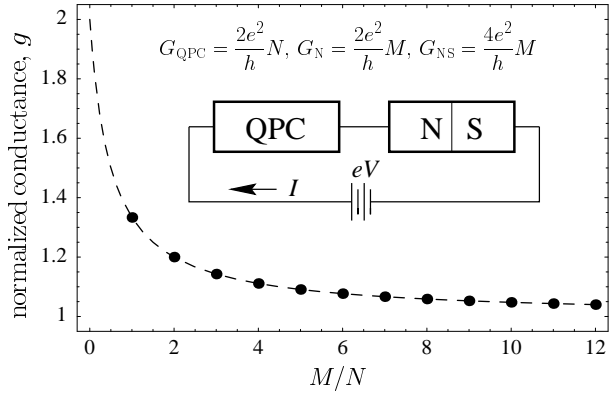, width=0.99\columnwidth}
\end{center}
\caption[]{Normalized conductance $g\equiv G_{\scriptscriptstyle\rm
    QPC-NS}/G_{\scriptscriptstyle\rm QPC-N}$ of a QPC NS junction with
  sequential tunneling through the ideal QPC and the ideal NS
  interface as a function of the ratio $M/N$ of propagating modes in
  the QPC and at the NS interface.}
\label{FIG6}
\end{figure}

For an ideal QPC and an ideal interface we have
$G_{\scriptscriptstyle\rm QPC}=\frac{2e^2}{h}N$,
$G_{\scriptscriptstyle\rm NS}=\frac{4e^2}{h}M$, and
$G_{\scriptscriptstyle\rm N}=\frac{2e^2}{h}M$, where $N$ is the number
of modes in the QPC and $M$ is the number of modes at the NS
interface. The corresponding normalized conductance is shown in Fig.
\ref{FIG6}.

The sequential tunneling behavior may provide an explanation for the
unexpectedly small conductance enhancement seen in the experiments of
Benistant \emph{et al.} \cite{BENISTANT85} on Ag-Pb interfaces with
injection of quasiparticles into an Ag crystal through a point
contact. The condition for the electronic transport to be incoherent
is that the distance between the point contact and the NS interface is
longer than the correlation length $L_c={\rm Min}(\ell_{\rm in},L_{\rm
  T})$, $\ell_{\rm in}$ being the inelastic scattering length and
$L_{\rm T}$ the Thouless length \cite{IMRY,BEENAKKER95}.  For the
ballistic device studied by Benistant {\it et al.} $L_c=L_{\rm
  T}=\hbar v_{\rm F}/k_{\rm B}T \sim 9\,{\rm \mu m}$ (at $T=1.2\,{\rm
  K}$) which is much shorter than the distance between the point
contact and the NS interface ($\sim 200\,{\rm \mu m}$). Lowering the
temperature will increase the correlation length and for sufficiently
low temperatures ($T \sim 0.05\,{\rm K}$) we expect a cross-over from
the sequential tunneling regime to the phase-coherent regime where the
Andreev mediated conductance enhancement should become observable.

\section{Discussion and conclusion}

For an ideal 2DEG-S interface with a QPC in the normal region, the
normalized conductance $g\equiv G_{\scriptscriptstyle\rm
  NS}/G_{\scriptscriptstyle\rm N}$ depends strongly on the position of
the Fermi level and only when the Fermi level corresponds to a
conductance plateau a doubling of the conductance is found. The
deviations from the factor-of-two enhancement, when the Fermi level
does not correspond to a plateau, can be significant and for a
particular example of the WNW constriction we find that the normalized
conductance can be suppressed to $g\sim 1.5$ in a system with only two
propagating modes in the constriction. In the presence of a barrier at
the 2DEG-S interface, the normalized conductance depends strongly on
the longitudinal quantization in the cavity set up by the QPC and the
barrier. Depending on the barrier strength, the length of the cavity
and the position of the Fermi level, this longitudinal quantization
may give rise to both constructive and destructive inferences in the
transmission and thus also in the conductance. Perhaps surprisingly,
the effect of the barrier is very much suppressed (compared to a
system without a QPC, see e.g. \cite{MORTENSEN99}) due to a very
strong back-scattering at the return of the quasiparticles to the
normal probe.  The localization of quasiparticles in the cavity gives
rise to an almost reflectionless tunneling through the barrier as it
is also found in systems with a diffusive normal region
\cite{BEENAKKER92,VANWEES92}.  The interferences due to localization
in the cavity will be smeared by a finite temperature and they are
also expected to be suppressed by a finite inelastic scattering length
compared to the length of the cavity \cite{SUN99}.

For the sequential tunneling regime we find that the conductance
enhancement vanishes as the number of modes at the interface becomes
much larger than the number of modes in the QPC.

Our calculations show that the S--matrix approach provides a powerful
alternative to the time-dependent Bogoliubov-de Gennes approach of De
Raedt \emph{et al.} \cite{RAEDT94} in detailed studies of the
conducting properties of nanoscale 2DEG-S devices. Even though the
back-focusing phenomenon of the Andreev reflection is robust against
changes in the geometry \cite{RAEDT94}, the electron-hole conversion
efficiency itself is not.

Finally, we stress that for a quantitative comparison to experimental
systems, it is crucial to take different Fermi wave vectors and
effective masses of the 2DEG and the superconductor into account
\cite{MORTENSEN99}.

\section*{Acknowledgements}
We would like to thank C.W.J. Beenakker, M. Brandbyge, J.B. Hansen,
and H.M. R\o nnow for useful discussions. NAM acknowledges financial
support by the Nordic Academy for Advanced Study (NorFA) and HS
acknowledges support by the European Community.

\widetext

\end{document}